  \documentclass[preprint, superscriptaddress,showkeys,showpacs,
 preprintnumbers,amsmath,amssymb,aps,floatfix]{revtex4-1}
\usepackage[USenglish]{babel}
\usepackage{graphicx}
\usepackage{epsfig}
\usepackage{amsfonts}
\usepackage{amsmath}
\usepackage{amssymb}
\usepackage[sort&compress]{natbib}
\usepackage{dcolumn}
\usepackage{multirow}
\usepackage{bigstrut}
\usepackage{mathrsfs}
\usepackage{type1cm}
\usepackage{placeins}
\usepackage{rotating}
\usepackage{slashed}

\begin{document}
\newcommand{\HRule}{\rule{\linewidth}{0.5mm}}
\newcommand{\cev}[1]{\reflectbox{\ensuremath{\vec{\reflectbox{\ensuremath{#1}}}}}}

\title{Exploration of Lorentz violation in neutral-kaon decay}

\author{K. K. Vos}
\thanks{Corresponding author} \email{k.k.vos@rug.nl} 

\affiliation{KVI, University of Groningen, Zernikelaan 25, 9747 AA Groningen, The Netherlands}

\author{J. P. Noordmans}
\affiliation{KVI, University of Groningen, Zernikelaan 25, 9747 AA Groningen, The Netherlands}
\author{H. W. Wilschut}
\affiliation{KVI, University of Groningen, Zernikelaan 25, 9747 AA Groningen, The Netherlands}
\author{R. G. E. Timmermans}
\affiliation{KVI, University of Groningen, Zernikelaan 25, 9747 AA Groningen, The Netherlands}

\date{\today}

\begin{abstract}
The KLOE collaboration recently reported bounds on the directional dependence of the lifetime of the
short-lived neutral kaon $K^0_S$ with respect to the dipole anisotropy of the cosmic microwave background. We interpret their results
in an effective field theory framework developed to probe the violation of Lorentz invariance in the weak
interaction and previously applied to semileptonic processes, in particular $\beta$ decay. In this approach
a general Lorentz-violating tensor $\chi^{\mu\nu}$ is added to the standard propagator of the $W$ boson.
We perform an exploratory study of the prospects to search for Lorentz violation in nonleptonic decays.
For the kaon, we find that the sensitivity to Lorentz violation is limited by the velocity of the kaons and
by the extent to which hadronic effects can be calculated. In a simple model we derive the $K^0_S$
decay rate and calculate the asymmetry for the lifetime. Using the KLOE data, limits on the values
of $\chi^{\mu\nu}$ are determined.
\end{abstract}

\maketitle

\section{Introduction}
The KLOE collaboration recently reported a precision measurement of the lifetime of the short-lived neutral kaon $K^0_S$ \cite{KLOElife, dirvio}. In addition, a search was made for the dependence of the lifetime on the direction of the $K^0_S$ with respect to the dipole anisotropy of the Cosmic Microwave Background (CMB). The asymmetry in the lifetime was measured to be less than about $10^{-3}$. In this paper we interpret the KLOE findings in a general effective field theory framework developed in Ref. \cite{Jacob1} to study the possibility of Lorentz violation in the weak interaction, in particular in neutron and allowed nuclear $\beta$ decay.  A broad class of Lorentz-violating effects was considered, in which the standard low-energy propagator of the $W$-boson is modified to
\begin{equation}\label{eq:Wboson}
\left\langle W^{\mu+}(q)W^{\nu-}(-q)\right\rangle = -i\,(g^{\mu\nu}+\chi^{\mu\nu})/M_W^2 \ ,
\end{equation}
where the complex tensor $\chi^{\mu\nu}$ describes the effects of Lorentz violation in the weak interaction. In particular, such a tensor arises in the Standard Model Extension (SME) of Kosteleck\'y and collaborators \cite{kossme,kospara}, 
an effective field theory describing Lorentz violation at low energies. The new Lorentz-violating terms could originate from spontaneous Lorentz violation in, for instance, unifying theories of quantum gravity~\cite{Lib09}. 
In Ref.~\cite{Jacob2}, this theoretical framework was extended to forbidden $\beta$ decay, leading to strong bounds on $\chi^{\mu\nu}$. In addition, bounds were derived recently for allowed $\beta$ decay \cite{Hans, Stefan} and in pion decay \cite{Alt13}.

Taking the KLOE measurement as an example to study Lorentz violation of the form of Eq.~(\ref{eq:Wboson}) in nonleptonic decays, we explore to which extent nonleptonic decays can compete with the bounds from semileptonic decays.
The contributions of QCD (gluon) corrections in nonleptonic decays are not fully understood theoretically. It has been claimed that QCD effects cause an enhancement of the $\Delta I=1/2$ decay modes and that this is at least partly due to so-called ``penguin diagrams.'' On the other hand, recent lattice QCD results \cite{Boy13} shed doubt on the importance of penguin diagrams. Since in this work we aim to explore the generic features of Lorentz violation in nonleptonic decays, it is beyond our scope to derive the full effective weak Hamiltonian that includes Lorentz violation. We calculate the contribution of tree-level $W$ exchange and show how this constrains $\chi^{\mu\nu}$. We find that the asymmetry is proportional to $\gamma^2$, where $\gamma$ is the Lorentz boost factor, favoring experiments with high-velocity kaons. In the Appendix, we demonstrate that the penguin diagram does not contribute to the Lorentz-violating part of the $K^0_S$ decay rate. Therefore, the sensitivity of the $K^0_S$ lifetime to Lorentz violation is further reduced by an amount which depends on the relative contribution of the penguin diagram.

\section{Nonleptonic neutral-kaon decay}
First, we briefly review the calculation of the $K^0_S$ decay rate into two pions in the SM \cite{Okun} and we discuss the $\Delta I =1/2$ rule. The neutral-kaon system is described not by the mass eigenstates, but by the CP eigenstates
\begin{equation}\label{eq:Kshort}
K^0_1 \equiv  \frac{K^0+\bar{K}^0}{\sqrt{2}} \sim K^0_S\ \ \ \ \ \textrm{and} \ \ \ \ \ K^0_2\equiv \frac{K^0 - \bar{K}^0}{\sqrt{2}}\sim K^0_L \ .
\end{equation}
The short-lived and long-lived kaons, $K^0_S$ and $K^0_L$, are approximately equal to the CP eigenstates $K^0_1$ and $K^0_2$. We neglect the small effect of CP violation and set $K^0_S\equiv K^0_1$. The short-lived kaon decays into two pions, $K^0_S \rightarrow \pi^+ \pi^-$, $\pi^0\pi^0$, a strangeness-changing transition with $\Delta S=1$.
The two pions in the final state can have isospin $I=0$, a $\Delta I=1/2$ transition, and $I=2$, a $\Delta I=3/2$ transition. Experimentally it is found that the first transition is enhanced compared to the latter. The origin of this enhancement is an open standing problem and is referred to as the $\Delta I =1/2$ rule. If this were an exact rule only the $\Delta I =1/2$ transition would be allowed in the SM, the ratio of the decay rates of the two final states would be
 \begin{equation}\label{eq:ration}
 \frac{W(K^0_S \rightarrow \pi^+\pi^-)}{W(K^0_S \rightarrow \pi^0\pi^0)} = 2 \ .
 \end{equation}
From experiments this ratio is found to be 2.26, implying a small contribution from the $\Delta I = 3/2$ transition. To quantify the $\Delta I =1/2$ enhancement, we can express the $K_S^0$ decay amplitudes in terms of $A_0$, the amplitude for the $I=0$ final state, and $A_2$, the amplitude for the $I=2$ final state. Using the experimental value for the ratio in Eq.~\eqref{eq:ration}, we find 
\begin{equation}
\frac{\textrm{Re}\:A_2}{\textrm{Re}\;A_0}\simeq 4.4 \% \ ,
\end{equation} 
which shows the large enhancement of the $\Delta I = 1/2 $ transition.

In the SM, nonleptonic $\Delta S =1$ decays are usually described theoretically by an effective interaction, which is obtained by dressing the weak Hamiltonian with hard-gluon corrections. These corrections change the coefficients and the operator structure of the Hamiltonian.
The hard-gluon corrections then also induce a $\Delta I =3/2$ operator. Calculations with this effective Hamiltonian show an enhancement of the $\Delta I =1/2$ transition, though insufficient to explain the experimental data.
The effective Hamiltonian contains six operators and their Wilson coefficients \cite{leffart}. Schematically, 
\begin{equation}
\mathcal{H}_{\textrm{eff}} \sim \frac{4G_F}{2\sqrt{2}} \cos\theta_C\sin\theta_C \sum_{i=1}^6 c_i \mathcal{O}_i \ , 
\end{equation}
where $G_F$ is the Fermi constant, $\theta_C$ is the Cabibbo angle, and $c_i$ are the Wilson coefficients of the operators $\mathcal{O}_i$. They can be found in Ref. \cite{leffart}. The dominant contributions to the $\Delta I = 1/2$ transition are given by $\mathcal{O}_1$ and $\mathcal{O}_5$,
\begin{subequations}
\begin{align}
\mathcal{O}_1 {}& = \bar{d}_L \gamma_{\mu}u_L \bar{u}_L \gamma^{\mu} s_L - \bar{u}_L \gamma_{\mu}u_L \bar{d}_L \gamma^{\mu} s_L \ , \\
\mathcal{O}_5 {}& = \bar{d}_L \gamma_{\mu} t^a s_L \left(\bar{q}_R \gamma^{\mu} t^a q_R\right) \ ,
\end{align}
\end{subequations}
where the subscript $L,R$ denotes the chirality of the quark and $t^a$ are the Gell-Mann matrices. 
Operator $\mathcal{O}_1$ arises from hard-gluon corrections to the tree-level diagram. 
The running of QCD logarithms gives a large coefficient $c_1$.

QCD enhancements also requires the inclusion of the so-called ``penguin diagram.'' The penguin diagram can be written as an effective interaction that generates $\mathcal{O}_5$, where gluon exchange makes it possible to couple to right-handed quarks. This results in an enhancement of the hadronic matrix elements. 

The combination of $\mathcal{O}_1$ and $\mathcal{O}_5$ gives the largest contribution to the decay rate, although even optimistic estimates of the matrix elements still find an amplitude that is a factor 5 too small compared to experimental data \cite{donpen}. 

In the SM, all operators of the effective Hamiltonian can be related to the form of $\mathcal{O}_1$ by Fierz transformations and Dirac algebra. The amplitude for $K^0$ decay into $\pi^+\pi^-$ in the SM can thus be written as
\begin{eqnarray} \label{eq:smkaon}
\left\langle\pi^+\pi^-|\mathcal{H}_{\textrm{eff}}|K^0\right\rangle {}& = & C_{\textrm{SM}} \left\langle \pi^+|\bar{u}_L\gamma^{\mu}d_L|0\right\rangle \left\langle \pi^-|\bar{s}_L\gamma_{\mu}u_L|K^0\right\rangle \nonumber \\
& = & \frac{1}{4} C_{\textrm{SM}} f_{\pi} (p_+\cdot p_K+ p_+\cdot p_-)
   = \frac{1}{4} C_{\textrm{SM}} f_{\pi} (m_K^2-m_{\pi}^2) \ , 
\end{eqnarray} 
where $p_K$, $p_+$, and $p_-$ are the $K^0$, $\pi^+$, and $\pi^-$ momenta, respectively, and $f_\pi\simeq 0.95 m_{\pi}$ is the pion decay constant. To find the second equality we use that the $K^0-\pi^-$ matrix element is proportional to $f_+(p_K+p_-)^\mu+ f_-(p_K-p_-)^{\mu}$, where the latter term can be neglected, since experiments give $f_-\ll f_+ \sim 1$. 
The coefficient $C_{\textrm{SM}}$ contains factors from Fierz transformations and Dirac algebra. The matrix element for $\bar{K}^0$ decay is the complex conjugate of the matrix element for $K^0$ decay, with the same $C_{\textrm{SM}}$. 

When we include Lorentz violation, we can no longer separate the amplitude into two matrix elements, as in Eq.~\eqref{eq:smkaon}, which are contracted with the $W$ boson propagator. Mixing between the different operators and new structures from Fierz transformations complicate the Lorentz-violating case even further. For a complete analysis the effective Hamiltonian with Lorentz violation should be calculated, this is however beyond the scope of our present work since we only wish to explore the possibilities for testing Lorentz-violation in nonleptonic decays. We shall instead use a theoretical model in which we consider tree-level $W$ exchange. In the Appendix we discuss the Lorentz-violating contribution to operator $\mathcal{O}_5$.    


\section{Theoretical model}
We will derive the decay rate of $K^0_S$ into $\pi^+\pi^-$ in a tree-level $W$-exchange model. 
For the Lorentz-violating amplitude of $K^0$ decay the modified $W$-boson propagator from Eq.~(\ref{eq:Wboson}) is inserted between the matrix elements in Eq.~(\ref{eq:smkaon}), 
\begin{equation}
\left\langle\pi^+\pi^-|\mathcal{H}|K^0\right\rangle = 2\sqrt{2} G_F \cos\theta_C\sin\theta_C \left\langle \pi^+|\bar{u}_L\gamma_{\mu}d_L|0\right\rangle (g^{\mu\nu}+\chi^{\mu\nu*})\left\langle \pi^-|\bar{s}_L\gamma_{\nu}u_L|K^0\right\rangle \ , 
\end{equation}
where the Hamiltonian only contains the tree-level operator. 
The differential decay rate of $K^0_S$ in the laboratory frame is given by
\begin{eqnarray} \label{eq:decayrate}
\frac{dW}{dE_+} & = &\frac{8 G_F^2 \cos\theta_C^2\sin\theta_C^2 f_{\pi}^2}{128\pi |\vec{p}_K|E_K} (m_K^2-m_{\pi}^2) \bigg[(m_K^2-m_{\pi}^2)  \nonumber \\
&&  + \chi_r^{00}(E_K^2+2E_KE_+-2E_+^2)  - (\chi_r^{i0}+\chi_r^{0i})p_K^i\left(E_K+E_+\right) + \chi_r^{ij} p_K^ip_K^j   \nonumber \\
&&   +\left[-(\chi_r^{i0} +\chi_r^{0i})(E_K-2E_+) p^i_K  +2\chi_r^{ij}  p_K^ip_K^j\right] \frac{2 E_K E_+ - m_K^2}{2|\vec{p}_K|^2} \nonumber \\
&&  - \left(\chi_r^{00}-\chi_r^{ij} \frac{p_K^i p_K^j}{|\vec{p}_K|^2} \right)(E_+^2-m_{\pi}^2) - \left(3\chi_r^{ij}\frac{p_K^i p_K^j}{|\vec{p}_K|^2}-\chi_r^{00}\right) \left(\frac{2 E_K E_+ - m_K^2}{2|\vec{p}_K|} \right)^2 \bigg] \ ,
\end{eqnarray}
where $\chi_r^{\mu\nu}$ is the real component of $\chi^{\mu\nu}$, 
we sum over repeated indices, and Latin indices run over $1,2,3$. 
The total decay rate is found by integrating over the pion energy between the boundaries
\begin{equation}
E_+ = \frac{1}{2} E_K \pm \frac{1}{2}|\vec{p}_K| \sqrt{1-\frac{4 m_{\pi}^2}{m_K^2}} \ .
\end{equation}
We find
\begin{eqnarray}\label{eq:fulldecay}
W & = &\frac{8 G_F^2 \cos\theta_C^2\sin\theta_C^2  f_{\pi}^2}{128\pi E_K} (m_K^2-m_{\pi}^2) \sqrt{1-\frac{4m_{\pi}^2}{m_K^2}} \nonumber \\
&& \times\left[(m_K^2-m_{\pi}^2) + \frac{4}{3} \chi^{\mu\nu}_r (p_K)_{\mu} (p_K)_{\nu}\left(1+\frac{m_{\pi}^2}{2m_K^2} \right)\right] \ .
\end{eqnarray}

In general, the tensor $\chi^{\mu\nu}$ in Eq.~(\ref{eq:Wboson}) can depend on the $W$-boson momentum $q$, where for $K^0$ decay $q=p_+$ and for $\bar{K}^0$ decay $q=p_-$. A momentum-dependent $\chi^{\mu\nu}$ complicates the integrals over the angle between the directions of the $K^0_S$ and the $\pi^+$, as discussed in Appendix B of Ref. \cite{Jacob1}. Here, we have restricted ourselves to a momentum-independent $\chi^{\mu\nu}$, because momentum-dependent parts are suppressed by powers of the $W$-boson mass. This can be shown explicitly in the minimal SME (mSME), the subset of the SME that is renormalizable and only contains terms up to mass dimension four \cite{kossme}. In the mSME the W-boson propagator, in the unitarity gauge and to first order in Lorentz violation, reads \cite{Jacob1}
\begin{eqnarray}
\left\langle W^{\mu+}(q)W^{\nu-}(-q)\right\rangle & = & \frac{-i}{q^2-M_W^2} \left\{g^{\mu\nu} - \frac{q^{\mu}q^{\nu}}{M_W^2} + \frac{M_W^2}{q^2-M_W^2}(k_{\phi\phi}^{\mu\nu} +\frac{i}{2g}k_{\phi W }^{\mu\nu}) \right. \nonumber \\
 && \left. -\frac{1}{q^2-M_W^2}\left[2k_W^{\rho\mu\sigma\nu}q_{\rho}q_{\sigma} +q^{\mu}q_{\rho}(k_{\phi\phi}^{\rho\nu}+\frac{i}{2g}k_{\phi W}^{\rho\nu}) \right. \right. \nonumber \\
 && \left. \left. + q^{\nu}q_{\rho}(k_{\phi\phi}^{\rho\mu}+\frac{i}{2g}k_{\phi W}^{\rho\mu})\right]+\frac{k_{\phi\phi}^{\rho\sigma}q_{\rho}q_{\sigma}q^{\mu}q^{\nu}}{M_W^2(q^2-M_W^2)}   \right\} \ ,
\end{eqnarray}
where $k_{\phi\phi}$, $k_{\phi W}$ and $k_{W}$ are SME parameters \cite{kossme}, and $g$ is the SU(2) electroweak coupling constant. Comparing this to the low-energy propagator in Eq.~\eqref{eq:Wboson} and neglecting momentum-dependent terms one finds \cite{Jacob1}
\begin{equation}\label{eq:SME}
\chi^{\mu\nu} = -(k_{\phi\phi})^{\mu\nu}-\frac{i}{2g}(k_{\phi W})^{\mu\nu} \ .
\end{equation}

Following the discussion in Ref.~\cite{Jacob1} we remark that Eq.~\eqref{eq:SME} agrees with the low-energy limit for the massive photon propagator \cite{Cam12} and with Ref.~\cite{Alt12}. Furthermore, a Lorentz-violating correction to the quark-quark-$W$ vertex gives the same structure for the effective interaction as Eq.~(\ref{eq:Wboson}) gives, but is more involved due to corrections to external quark states~\cite{Jacob1}. 
The tensor $\chi^{\mu\nu}$ can be both CPT-odd and CPT-even, but when considering only momentum-independent terms it is CPT-even. 
Since we only consider momentum-independent modifications to the $W$-boson propagator, hermiticity of the Lagrangian implies that $\chi^{\mu\nu*} = \chi^{\nu\mu}$~\cite{Jacob1}.  

\section{Constraints on Lorentz violation from the KLOE data}
With the KLOE detector at DA$\Phi$NE, decay branching ratios of kaons \cite{vuskaon} were measured to determine the value of the element $V_{us}$ of the quark-mixing matrix. The $K^0_S$ mesons were created in the strong decay  $\phi\rightarrow K^0_L K^0_S$, where the long-lived $K^0_L$ is not detected. The $K^0_S$ lifetime was measured with high precision \cite{KLOElife}. The collaboration also measured the difference in the $K^0_S$ lifetime parallel ($\tau^+$) and lifetime antiparallel ($\tau^-$) to a direction fixed in space, with the asymmetry defined as
\begin{equation}
\mathcal{A} =\frac{\tau^+ - \tau^-}{\tau^++\tau^-} \ .
\end{equation}
The $K^0_S$ momenta in the laboratory frame were transformed event-by-event to galactic coordinates \cite{doctomaria} specified by $\left\{\ell, b\right\}$, where $\ell$ is the galactic longitude and $b$ is the galactic latitude.
\begin{table}[t]
\begin{tabular}{l|r} 
 \hline\hline $\left\{\ell,b\right\}$ & $\mathcal{A}_{\textrm{cone}}\times 10^3$  \\ \hline
 CMB0 = $\left\{264^\circ, 48^\circ\right\}$ & $-0.2\pm1.0$  \cite{KLOElife}\\
 CMB0 = $\left\{264^\circ, 48^\circ\right\}$ & $-0.13\pm0.4$  \cite{dirvio}\\
 CMB1 = $ \left\{174^\circ, 0^\circ\right\}$ & $0.2\pm1.0$  \cite{KLOElife}\\
 CMB2 =  $\left\{264^\circ, -42^\circ\right\}$ & $0.0\pm0.9$  \cite{KLOElife} \\ \hline\hline
\end{tabular}
\caption{Observed $K^0_S$ lifetime asymmetry \cite{KLOElife, dirvio}, where $\left\{\ell,b\right\}$ are the galactic coordinates. CMB0 is the direction of the dipole anisotropy in the CMB and CMB1 and CMB2 are two perpendicular directions. The errors are mainly statistical.}
\label{table:as}
\end{table}
The asymmetry was measured in three different directions in the CMB rest frame.  The first direction, $\left\{264^\circ, 48^\circ\right\}$, is the direction of the CMB dipole anisotropy. The directions labeled CMB1 and CMB2 are two perpendicular directions. Only events inside a cone of $30^\circ$ opening angle were used, resulting in a difference between the cone asymmetry and the asymmetry for one specific direction $\vec{n}$,
\begin{equation}
\mathcal{A}_{\textrm{cone}} \simeq  0.93\,\mathcal{A}_{\vec{n}} \ .
\end{equation}
The KLOE results for $\mathcal{A}_{\textrm{cone}}$ for the different directions are listed in Table \ref{table:as}.

In our framework, the $K^0_S$ lifetime asymmetry can be constructed from the decay rate in Eq.~(\ref{eq:fulldecay}).
The KLOE collaboration measured charged pions coming from $K^0_S$ decay in different directions, and derived from this the total decay rate. In the quoted asymmetry we thus need the total $K^0_S$ lifetime, which includes the decay into two neutral pions. We found that the neutral decay does not acquire additional Lorentz-violating contributions, and the ratio between the two main decay modes in Eq.~\eqref{eq:ration} is therefore not altered.
We find
\begin{eqnarray}\label{eq:asym}
\mathcal{A}_{\vec{n}} & = & \frac{ \frac{4}{3}+\frac{2}{3}\frac{m_{\pi}^2}{m_K^2}}{m_K^2-m_{\pi}^2}\,(\chi_r^{i0}+\chi_r^{0i})\,E_Kp_K^i \nonumber \\  & = & \frac{\frac{4}{3}+\frac{2}{3}\frac{m_{\pi}^2}{m_K^2}}{\left(1-\frac{m_{\pi}^2}{m_K^2}\right)} \,\gamma^2\,\chi^{i0}_S\,\beta_K^i \ ,
\label{An1}
\end{eqnarray}
where  $\chi^{i0}_S\equiv\chi_r^{i0}+\chi_r^{0i}$, and $\beta_K$ is the velocity of the $K^0_S$. Because the $K^0_L$ and $K^0_S$ originate from a $\phi$-meson created nearly at rest in $e^+e^-$ collisions, such that $\beta_{K}$=0.217 and $\gamma=1.02$, this gives
\begin{equation}
\mathcal{A}_{\vec{n}} = 0.34\,\chi^{i0}_S\,\hat{\beta}_K^i \ ,
\label{An2}
\end{equation}
where $\hat{\beta}_K$ is the direction of the $K^0_S$ velocity.

Several observations about this result should be made.
The asymmetry in Eq.~\eqref{eq:asym} shows a $\gamma^2$ enhancement, and a dependence on the real and symmetric part of $\chi^{\mu\nu}$ that transforms as a vector under rotations. This is a general result, {\em i.e.} the most advantageous way to measure Lorentz-violating effects in weak decays is from a fast-moving decaying particle. Only then can one compete with the results from forbidden $\beta$
decay~\cite{Jacob2}, which profited from the high statistics one can obtain with a high-intensity source.  
Considering the contribution of the ${\cal O}_5$ operator discussed in the Appendix, we find no dependence on
$\chi^{\mu\nu}$ when evaluating the dependence of the transition strength on the decay direction. Assuming that indeed the dominant contributions to the decay rate are from ${\cal O}_1$ and ${\cal O}_5$, 
the actual dependence on $\chi^{\mu\nu}$ in Eqs.~({\ref{An1}) and ({\ref{An2}) is reduced. The precise 
reduction depends on the relative amplitudes of the two operators and its evaluation is complicated by theoretical uncertainties in the hadronic effects. In this respect, semileptonic decays are at this moment theoretically favorable for Lorentz-violation tests.

To see what type of limits one may obtain, we ignore these caveats. From the KLOE data, we can then put a 95\% confidence limit (C.L.) bound on $\chi_S^{i0}$ in the CMB direction of
\begin{subequations}
\begin{equation}\label{eq:bound}
|\chi^{\textrm{\tiny CMB0},0}_S| < 2.9 \times 10^{-3} \;(95\%\;\;\textrm{C.L.}) \ .
\end{equation}
For the other two directions we find
\begin{eqnarray}
|\chi^{\textrm{\tiny CMB1},0}_S| & < & 6.8 \times 10^{-3} \;(95\%\;\textrm{C.L.}) \ , \\
|\chi^{\textrm{\tiny CMB2},0}_S| & < & 5.5 \times 10^{-3} \;(95\%\;\textrm{C.L.}) \ .
\end{eqnarray}
\end{subequations}

For completeness and comparison between experiments, we transform the bounds from the KLOE asymmetries to the Sun-centered frame \cite{kospara}, in which $\hat{Z}$ is parallel to Earth's rotational axis, $\hat{X}$ points to the vernal equinox at time $t=0$, and $\hat{Y}$ completes the right-handed coordinate system. To evaluate the bounds in the Sun-centered frame we first transform the galactic coordinates $\left\{\ell,b\right\}$ to equatorial coordinates $(\alpha, \delta)$ via
\begin{subequations}
\begin{eqnarray} 
\delta & =  & \sin^{-1}\left[\cos {b} \cos (27.4^\circ)\sin(\ell-33^\circ)+\sin b\sin(27.4^\circ)\right] \ , \\
\alpha & = & \tan^{-1}\left[\frac{\cos b \cos(\ell-33^\circ)}{\sin b \cos(27.4^\circ)-\cos b \sin(27.4^\circ)\sin(\ell-33^\circ)} \right] +192.25^\circ \ ,
\end{eqnarray}
\end{subequations}
where $\alpha$ is the right-ascension and $\delta$ is the declination. The equatorial coordinates can then be transformed to the Sun-centered frame $\left\{T, X, Y, Z\right\} \equiv \left\{T,\vec{I}\,\right\}$ by using
$\vec{I} = (\cos\delta \cos\alpha, \cos\delta \sin\alpha,\sin\delta)$.
For the CMB directions this gives
\begin{subequations}
\begin{eqnarray}
\chi^{\textrm{\tiny CMB},0}_S & = & -0.97 X^{XT}_S + 0.22 X^{YT}_S -0.11 X^{ZT}_S \ , \\
\chi^{\textrm{\tiny CMB1},0}_S & = & 0.12 X^{XT}_S + 0.82 X^{YT}_S + 0.56 X^{ZT}_S \ , \\
\chi^{\textrm{\tiny CMB2},0}_S & = & 0.22 X^{XT}_S + 0.52 X^{YT}_S -0.82 X^{ZT}_S \ ,
\end{eqnarray}
\end{subequations}
where $X^{\mu\nu}_S\equiv X^{\mu\nu}_r+X^{\nu\mu}_r$ are the Lorentz-violating quantities in the Sun-centered frame. 
For the values in the Sun-centered frame we then find
\begin{subequations}
\begin{eqnarray}
 |X^{XT}_S| & < & 3.3 \times 10^{-3}  \;(95\%\;\textrm{C.L.}) \ , \\
 |X^{YT}_S| & < & 6.3 \times 10^{-3} \;(95\%\;\textrm{C.L.}) \ ,  \\
 |X^{ZT}_S| & < & 6.0 \times 10^{-3} \;(95\%\;\textrm{C.L.})  \ . 
\end{eqnarray}
\end{subequations}

\section{Summary and outlook}
In this paper, we explored the possibilities to test Lorentz violation in nonleptonic decays, taking the KLOE $K_S^0$ lifetime asymmetry measurement as an example. We used the framework developed in Ref.~\cite{Jacob1}, in which Lorentz violation in the weak interaction is studied by introducing a general Lorentz-violating tensor $\chi^{\mu\nu}$, which modifies the $W$-boson propagator. We discussed the difficulties concerning nonleptonic decays within the SM and restricted ourselves to a simplified model. We calculated the directional asymmetry of the $K^0_S$ lifetime, defined by the difference in lifetime between the $K^0_S$ decaying parallel and antiparallel to a specific direction in space. The KLOE collaboration measured this asymmetry with a precision of $10^{-3}$ in the direction defined by the CMB dipole. For this direction $\chi^{0i}_S$ is constrained to be less than $10^{-3}$. 
Our results put constraints on the SME parameters, for example $k_{\phi\phi}$, by relating our $\chi^{\mu\nu}$ to Eq.~\eqref{eq:SME}~\cite{Jacob1,Jacob2}. We find at $95 \%$ C.L.
\begin{subequations}\label{eq:SMElimits}
\begin{eqnarray}
 |(k_{\phi\phi})^{XT}_S| & < & 3.3 \times 10^{-3}  \ , \\
 |(k_{\phi\phi})^{YT}_S| & < & 6.3 \times 10^{-3} \ ,  \\
 |(k_{\phi\phi})^{ZT}_S| & < & 6.0 \times 10^{-3} \ . 
\end{eqnarray}
\end{subequations}

The long-standing problem of the $\Delta I =1/2$ rule shows the challenges of nonleptonic decays. In the usual effective Hamiltonian description the penguin diagram gives a large contribution, but we showed that the Lorentz-violating contribution to this penguin diagram vanishes. This would further reduce the sensitivity of the lifetime to Lorentz violation, which would worsen our bounds in Eq.~\eqref{eq:SMElimits}.
From a theoretical point of view, semileptonic and leptonic decays are at this point preferable for Lorentz-invariance tests. 
As far as the weak interaction is concerned bounds already exist from allowed \cite{Hans,Stefan} and forbidden \cite{Jacob2} $\beta$ decay and from pion decay \cite{Alt13}. Possibilities to complement and/or compete with these bounds lie in exploiting the $\gamma^2$ enhancement that occurs in asymmetries in experiments with high-energy hadrons.

\begin{acknowledgments}
We thank Alan Kosteleck\'y and Antonio De Santis for helpful discussions.
This research was supported by the Dutch Stichting voor Fundamenteel Onderzoek der Materie
(FOM) under Programmes 104 and 114 and project 08PR2636. \\
\end{acknowledgments}

\appendix
\section{Penguin diagram} 
The penguin diagram 
generates $\mathcal{O}_5$ and can be written as an effective vertex 
by integrating out the $W$ boson \cite{wisewitten}.
The Lorentz-violating (LV) operator is found by calculating this effective vertex with our modified $W$-boson propagator, 
\begin{eqnarray}\label{eq:penguin}
\mathcal{O}_5^{\textrm{LV}}& = & -\frac{1}{2} \bar{d}_L t^a \left[\chi^{\mu\nu}+\chi^{\nu\mu} + i \epsilon^{\alpha\beta\mu\nu}\chi_{\alpha\beta} \right] \gamma_{\nu} s_L \left( \bar{q}_R t^a \gamma_{\mu} q_R\right) \nonumber\\
&& -\frac{1}{2} \bar{s}_L t^a \left[\chi^{\mu\nu*}+\chi^{\nu\mu*} +  i \epsilon^{\alpha\beta\mu\nu}\chi^*_{\alpha\beta} \right] \gamma_{\nu} d_L \left( \bar{q}_R t^a \gamma_{\mu} q_R\right)\ .
\end{eqnarray}
To calculate the matrix elements we use the vacuum-saturation method, in which we insert a complete set of states between the initial and final state. Using Fierz transformations and Gell-Mann matrix algebra we can write Eq.~\eqref{eq:penguin} in a more convenient form. 
For the Lorentz-violating case these transformations are more involved than in the SM, as the Dirac matrices are no longer contracted with $g^{\mu\nu}$. The Fierz transformations now give additional Lorentz scalar and tensor structures. Due to parity constraints some of these structures do not contribute. 
We find
\begin{align}
\left\langle \pi^-\pi^+\left| \mathcal{O}^{\textrm{LV}}_5\right|\bar{K}^0\right\rangle {}& = 
-\frac{1}{2}\left[\chi^{\mu\nu}+\chi^{\nu\mu} + i \epsilon^{\alpha\beta\mu\nu}\chi_{\alpha\beta} \right] \left\langle \pi^-\pi^+\left| \bar{d}_L \gamma_{\nu} t^a s_L\bar{q}_R \gamma_{\mu}t^a q_R \right|\bar{K}^0\right\rangle  &&\nonumber \\
{}& = \frac{i}{8} B^{\mu\nu} \left\langle \pi^-\left|\bar{d}\gamma_5 u\right|0\right\rangle \left\langle \pi^+ \left|\bar{u} \sigma_{\mu\nu} s\right|\bar{K}^0\right\rangle \ ,
\end{align}
where
\begin{equation}
B^{\mu\nu} \equiv \chi^{\mu\nu} - \chi^{\nu\mu} - i \epsilon^{\alpha\beta\mu\nu}\chi_{\alpha\beta} \ ,
\end{equation}
and the matrix element $\left\langle \pi^-\left|\bar{d}\gamma_5u\right|0\right\rangle  = i f_{\pi} m_{\pi}^2/(m_u+m_d) $ \cite{donpen}, 
$\left\langle \pi^+(p)\left|\bar{u} \sigma_{\mu\nu} s\right|\bar{K}^0(k)\right\rangle = (p_{\mu} k_{\nu} - k_{\mu}p_{\nu}) 2f_T/(m_K+m_{\pi})$,
with $f_T=0.417$ \cite{tensor}. 
We can now calculate the amplitude for $K^0_S$ decay with $\mathcal{O}^{\textrm{LV}}_5$
\begin{equation}\label{eq:O5liv}
\left\langle \pi^-\pi^+\left| \mathcal{O}^{\textrm{LV}}_5\right|K^0_S\right\rangle = \frac{i}{\sqrt{2}}  C_{\textrm{LV}}\left(B_{\mu\nu}p_+^{\mu}p_K^{\nu}+\tilde{B}_{\mu\nu}p_-^{\mu}p_K^{\nu} \right) \ ,
\end{equation}
where $C_{\textrm{LV}}$ contains numerical Fierz and matrix element factors and $\tilde{B}_{\mu\nu}\equiv B_{\mu\nu}(\chi_{\mu\nu}\rightarrow \chi_{\mu\nu}^*)$.
The interference of the amplitude in Eq.~\eqref{eq:O5liv} and $\mathcal{M}_{\textrm{SM}}\equiv\left\langle\pi^+\pi^-|\mathcal{H}_{\textrm{eff}}|K^0_S\right\rangle=\sqrt{2}\left\langle\pi^+\pi^-|\mathcal{H}_{\textrm{eff}}|K^0\right\rangle$, given in Eq.~\eqref{eq:smkaon}, gives for the LV contributions to the differential decay rate
\begin{eqnarray}
\frac{dW^{\textrm{LV}}_5}{dE_+} & = & \frac{1}{16 \pi |\vec{p}_K|E_K} \biggl\{\frac{iC_{\textrm{LV}}}{\sqrt{2}} \mathcal{M}_{\textrm{SM}} \biggl[ (B_{0\nu}-B_{0\nu}^*-\tilde{B}_{0\nu}+\tilde{B}^*_{0\nu})E_+ p_K^{\nu} \nonumber \\
&& + (B_{i\nu}-B_{i\nu}^*-\tilde{B}_{i\nu}+\tilde{B}_{i\nu}^*)\hat{p}_K^i p_K^{\nu} \frac{2E_KE_+-m_K^2}{2|\vec{p}_K|}+(\tilde{B}_{\mu\nu}-\tilde{B}^*_{\mu\nu})p_K^{\mu}p_K^{\nu}\biggr]\biggr\}.
\end{eqnarray}
Performing the integration over $E_+$, we find that the contribution to the total decay rate of $\mathcal{O}_5^{\textrm{LV}}$ vanishes. This is anticipated since $B_{\mu\nu}$ is antisymmetric, while the $K^0_S$ four-momentum is the only non-LV variable the decay can depend on. The decay rate, which is observer Lorentz invariant, can thus only depend on $B_{\mu\nu} p_K^{\mu} p_K^\nu = 0$.


\begin{thebibliography}{99}

\bibitem{KLOElife}
F. Ambrosino {\it et al.} (KLOE Collaboration), Eur. Phys. J. C {\bf 71}, 1604 (2011).
  
\bibitem{dirvio}
A. De Angelis, M. De Maria, M. Antonelli, and M. Dreucci,
Nuovo Cim. {\bf C034N3}, 323 (2011).

\bibitem{Jacob1}
J. P. Noordmans, H. W. Wilschut, and R. G. E. Timmermans, 
Phys. Rev. C {\bf 87}, 055502 (2013).

\bibitem{kossme}
D. Colladay and V. A. Kosteleck\'y, Phys. Rev. D {\bf 55}, 6760 (1997);
Phys. Rev. D {\bf 58}, 116002 (1998).

\bibitem{kospara}
V. A. Kosteleck\'y and N. Russell, Rev. Mod. Phys. {\bf 83}, 11 (2011);
 {\tt arXiv:0801.0287[hep-ph]}.
 
\bibitem{Lib09} S. Liberati and L. Maccione, Annu. Rev. Nucl. Part. Sci. {\bf 59}, 245 (2009);
S. Liberati, Class. Quant. Grav. {\bf 30}, 133001 (2013).

\bibitem{Jacob2}
J. P. Noordmans, H. W. Wilschut, and R. G. E. Timmermans,
Phys. Rev. Lett. {\bf 111}, 171601 (2013).


\bibitem{Hans}
H. W. Wilschut {\it et al.}, Ann. Phys. (Berlin) {\bf 525}, 653 (2013).

\bibitem{Stefan}
S. E. M\"uller {\it et al.}, Phys. Rev. D {\bf 88}, 071901(R) (2013).

\bibitem{Alt13} B. Altschul, Phys. Rev. D {\bf 88}, 076015 (2013).

\bibitem{Boy13} P. A. Boyle {\it et al.} (RBC and UKQCD Collaborations),
Phys. Rev. Lett. {\bf 110}, 152001 (2013).

\bibitem{Okun}
L. B. Okun, \textit{Leptons and Quarks} (North Holland, 1985).
\bibitem{leffart}
M. Shifman, A. Vainshtein, and V. Zakharov, Nucl. Phys. {\bf B120}, 316 (1977).

\bibitem{donpen}
J. F. Donoghue, Phys. Rev. D {\bf 30}, 1499 (1984).

\bibitem{Cam12}
M. Cambiaso, R. Lehnert, and R. Potting,
Phys. Rev. D {\bf 85}, 085023, (2012)

\bibitem{Alt12}
B. Altschul,
Phys. Rev. D {\bf 86}, 045008, (2012)




\bibitem{vuskaon}
F. Ambrosino {\it et al.} (KLOE Collaboration), JHEP {\bf 0804}, 059 (2008).

\bibitem{doctomaria}
M. De Maria, Ph.D. thesis, University of Udine (2010).


\bibitem{wisewitten}
M. B. Wise and E. Witten, Phys. Rev. D {\bf{20}}, 1216 (1979).

\bibitem{tensor}
I. Baum {\it et al.}, Phys. Rev. D {\bf{84}}, 074503 (2011).  

\end{thebibliography}
\end{document}